\documentclass[12pt,preprint]{aastex}
\usepackage{color}

\slugcomment{accepted for publication in ApJ}

\shorttitle{Dust formation in the ejecta of SN 2006jc}
\shortauthors{Nozawa et al.}

\begin{document}

\title{Early Formation of Dust in the Ejecta of Type Ib SN 2006jc and
Temperature and Mass of the Dust}

\author{Takaya Nozawa,\altaffilmark{1} Takashi Kozasa,\altaffilmark{1}
Nozomu Tominaga,\altaffilmark{2,3} Itsuki Sakon,\altaffilmark{3} \\
Masaomi Tanaka,\altaffilmark{3} Tomoharu Suzuki,\altaffilmark{3}
Ken'ichi Nomoto\altaffilmark{3,4,5} Keiichi Maeda,\altaffilmark{5,6} \\
Hideyuki Umeda,\altaffilmark{3} Marco Limongi,\altaffilmark{7,8} and
Takashi Onaka\altaffilmark{3}}

\altaffiltext{1}{Department of Cosmosciences, Graduate School 
of Science, Hokkaido University, Sapporo 060-0810, Japan; 
tnozawa@mail.sci.hokudai.ac.jp}
\altaffiltext{2}{Optical and Infrared Astronomy Division, National
Astronomical Observatory, 2-21-1 Osawa, Mitaka, Tokyo 181-8588, Japan}
\altaffiltext{3}{Department of Astronomy, School of Science, University
of Tokyo, Bunkyo-ku, Tokyo 113-0033, Japan}
\altaffiltext{4}{Research Center for the Early Universe, School of
Science, University of Tokyo, Bunkyo-ku, Tokyo 113-0033, Japan}
\altaffiltext{5}{Institute for the Physics and Mathematics of the
Universe, University of Tokyo, Kashiwa, Chiba 277-8568, Japan}
\altaffiltext{6}{Max-Planck-Institut f\"ur Astrophysik,
Karl-Schwarzschild Strasse 1, 85741 Garching, Germany}
\altaffiltext{7}{Istituto Nazionale di Astrofisica-Osservatorio 
Astronomico di Roma, Via Frascati 33, I-00040, Monteporzio Catone, Italy}
\altaffiltext{8}{Center for Stellar and Planetary Astrophysics,
School of Mathematical Sciences, P.O. Box, 28M, Monash University, 
Victoria 3800, Australia}

\begin{abstract}
SN 2006jc is a peculiar supernova (SN), in which the formation of dust 
has been confirmed at an early epoch of $\sim$50 days after the
explosion.
We investigate the possibility of such an earlier formation of dust
grains in the expanding ejecta of SN 2006jc, applying the Type Ib SN
model that is developed to reproduce the observed light curve.
We find that the rapid decrease of the gas temperature in SN 2006jc 
enables the condensation of C grains in the C-rich layer at 40--60 days 
after the explosion, which is followed by the condensation of silicate 
and oxide grains until $\sim$200 days.
The average radius of each grain species is confined to be less than
0.01 $\micron$ due to the low gas density at the condensation time. 
The calculated total dust mass reaches $\simeq$1.5 $M_{\odot}$, of 
which C dust shares $0.7$ $M_{\odot}$. 
On the other hand, based on the calculated dust temperature, we show 
that the dust species and mass evaluated to reproduce the spectral 
energy distribution observed by {\it AKARI} and {\it MAGNUM} at day 200
are different from those obtained by the dust formation calculations;
the dust species contributing to the observed flux are hot C and FeS 
grains with masses of $5.6 \times 10^{-4}$ $M_{\odot}$ and 
$2.0 \times 10^{-3}$ $M_{\odot}$, respectively, though we cannot defy 
the presence of a large amount of cold dust such as silicate and oxide 
grains up to $0.5$ $M_{\odot}$.
One of the physical processes responsible for the difference between 
calculated and evaluated masses of C and FeS grains could be considered 
to be the destruction of small-sized clusters by energetic photons and
electrons prevailing within the ejecta at the earlier epoch.

\end{abstract}

\keywords{dust, extinction --- supernovae: general 
--- supernovae: individual (SN 2006jc)}

\section{Introduction}

Supernova (SN) 2006jc is a peculiar Type Ib SN (SN Ib) in which the 
formation of dust has been observationally confirmed at a very early 
time of $\sim$50 days after the peak brightness (Smith et al. 2008), 
which is more than a few hundred days earlier than those for the 
dust-forming SNe observed so far.
Evidence for dust formation in SN 2006jc comes from an increase of the 
red to near-infrared (NIR) continuum (Di Carlo et al. 2007) and 
simultaneous emergence of blueshifted narrow He I emission lines 
(Smith et al. 2008).
Smith et al. (2008) have proposed that the detected dust does not 
condense in the ejecta but forms in the dense postshock gas swept up by 
the forward and reverse shocks.
Although X-ray observations of {\it SWIFT} and {\it Chandra} (Immler et 
al. 2008) suggested the presence of the dense circumstellar (CS) shell 
ejected by the outburst similar to those seen in luminous blue variables 
(LBVs) two years prior to the explosion (Nakano et al. 2006; Pastorello 
et al. 
2007), the postshock CS gas density estimated from the X-ray light curve 
(Tominaga et al. 2007) is not high enough for dust grains to condense. 
Thus, it is reasonable to consider that the appearance of dust toward SN 
2006jc should be the outcome of ongoing dust formation in the expanding 
ejecta of the SN.

Several pieces of evidence for dust formation in the SN ejecta have been
reported for Type II SNe (SNe II):
SN 1987A (Lucy et al. 1989; Whitelock et al. 1989; Meikle et al. 1993;
Wooden et al. 1993; Colgan et al. 1994), 
SN 1998S (Gerardy et al. 2000; Pozzo et al. 2004),
SN 1999em (Elmhamdi et al. 2003), 
and SN 2003gd (Sugerman et al. 2006; Meikle et al. 2007).
For these SNe II except for SN 1998S, the onset of dust formation is 
estimated to be later than 400 days after explosions.
Type IIn SN 1998S with the relatively low-mass hydrogen envelope (e.g., 
Liu et al. 2000; Fransson et al. 2005) exhibits signatures of dust 
condensation around 230 days (Gerardy 2000).
In addition, SN 1990I is the first SN Ib signifying the ongoing dust 
formation, in which dust formation is observed at $\sim$ 230 days 
(Elmhamdi et al. 2004).
On the other hand, the signature of dust formation in SN Ic has not 
been recorded so far.
Although theoretical studies have shown feasibility of dust formation in 
SNe II (Kozasa et al. 1989, 1991; Todini \& Ferrara 2001; Nozawa et al. 
2003), the possibility of dust formation in SNe Ib/c has never been
explored to date.
Hence, it is an important subject to be pursued how the formation 
process of dust in the ejecta depends on the type of SNe.

Since the progenitor stars of SNe Ib/c have lost most of the 
hydrogen/helium envelopes before the explosion, their ejected masses are
smaller, and the expansion velocities are significantly higher than SNe
II.
This leads to a lower density of gas in the ejecta, and the gas 
temperature drops down more quickly than those in typical SNe II.
This allows us to expect earlier dust formation in SNe Ib/c than SNe II.
On the other hand, the lower gas density may even result in unsuccessful 
dust formation.
Furthermore, in the early epoch, energetic photons and electrons
generated by the Compton degradation of $\gamma$-rays from the decay of 
radioactive elements such as $^{56}$Ni and $^{56}$Co prevail throughout 
the ejecta, and may affect the formation process of dust grains.
SN 2006jc is an ideal laboratory for examining the process of dust 
formation in SNe Ib because a copious amount of observational data
enable us to compare with theoretical models.
In this paper, we investigate, for the first time, the possibility of
dust formation in the expanding ejecta of SN Ib, based on the SN model 
that can well reproduce the light curve of SN 2006jc (Tominaga et al. 
2007).

The paper is organized as follows.
In \S~2 we summarize the observational evidence for dust formation 
in SN 2006jc and its interpretation, which are compared with the
calculations of dust formation in the later sections.
In \S~3 we investigate the dust formation in the ejecta and show that it 
is possible for dust grains to condense in the ejecta of SN 2006jc at
very early times of about 50 days after explosion.
In \S~4 we calculate the temperature of possible condensates obtained in 
\S~3 and evaluate the amount of dust grains contributing to the spectral 
energy distribution observed with {\it AKARI} and {\it MAGNUM} at day
200. 
Formation processes of dust grains in SN 2006jc are discussed in \S~5, 
focusing on the interpretation of the difference between dust mass 
obtained by the dust formation calculation and that evaluated from the 
spectral fitting.
The summary is presented in \S~6.
In this paper, we assume that the explosion of SN 2006jc has occurred at 
15 days before its discovery.

\section{Observations of Dust Formation in SN 2006jc and Its
 Interpretation}

\subsection{Evidence for dust formation}

The first verification of the presence of dust in SN 2006jc comes from 
the re-brightening at {\it J}, {\it H}, and {\it K} bands around day 50 
(Arkharov et al. 2006; Di Carlo et al. 2007).
In addition, Di Carlo et al. (2007) have noted a concomitant steep drop of
the light curve in the optical bands starting from 70 days, suggestive of 
strong extinction by the newly formed dust.
It is also shown that this rapid decline of the optical light curve 
continues up to $\sim$120 days (Tominaga et al. 2007; Anupama et al. 2008; 
Kawabata et al. 2008; Mattila et al. 2008).

Definitive evidence for dust formation has been reported by Smith et al. 
(2008); the asymmetry of narrow He I emission lines increased 
concurrently with increasing the red to NIR continuum between 65 and 120 
days.
The increasing red/NIR continuum is interpreted as the thermal radiation 
from the newly formed carbonaceous dust with temperature of $\sim$1600
K, considering its high temperature.
At day 140, the excess of this continuum disappeared, which indicates
that the dust grains were completely destroyed or the dust temperature 
became so low that its thermal emission was shifted to longer 
wavelengths.
The increasing asymmetry of narrow He I emission lines tracing the shocked 
CS gas was caused by the dust obscuration of the redshifted side, whereas 
broad emission lines tracing the SN ejecta such as Ca II and O I faded 
entirely.
The blueshifts of He I emission lines persisted up to day 180 (Mattila et 
al. 2008), asserting that the newly formed carbon grains were not 
destroyed but cooled down, which is supported by the NIR observations with 
{\it AKARI} (Sakon et al. 2007) and {\it MAGNUM} (Minezaki et al. 2007) 
showing that the temperature of the carbon grains dropped down to 
$\sim$800 K at day 215.

The time evolution of the optical light curve and the He I emission line 
profile during 50--140 days strongly suggests the ongoing formation of 
dust in SN 2006jc and reliably rules out as origins of the red/NIR 
continuum the scenarios involving pre-existing dust in the CS medium 
(CSM), such as the IR light-echo (Smith et al. 2008, Mattila et al. 2008) 
and the heating of the CS dust through the collision with the forward shock.
However, as suggested by Sakon et al. (2007) and Mattila et al (2008), 
the mid-IR spectrum observed at later epoch ($\ga$ 200 days) can be 
attributed to the low-temperature CS dust, which is discussed in \S~4.2.

\subsection{The site of dust formation}

Smith et al. (2008) have supposed that dust grains are formed in the
dense shell swept up by shocks but not in the freely expanding ejecta, 
based on (1) the geometrical considerations for explaining the blueshifts 
of narrow He I lines and the overall fading of broad Ca II and O I lines 
and (2) the time evolution of dust temperature that remains nearly 
constant for about a month and decreases after then; 
they considered that dust can continuously form at temperature of
$\sim$1600 K in the rapidly cooling regions behind the forward shock 
rather than the reverse shock, in the course of the passage of the forward 
shock through the dense CS shell created by the LBV-like eruption two 
years prior to the SN explosion.
This scenario is supported by a flux ratio of He I 
$\lambda7065$/$\lambda5876$ close to unity and the He II $\lambda4686$ 
emission feature appeared during the putative time of dust formation.
The detection of these He lines are expected to indicate the presence of 
the postshock gas of $\sim$$3 \times 10^{-14}$ g cm$^{-3}$ that is 
comparable to the critical density for carbon grain condensation (e.g., 
Clayton 1979), although Mattila et al. (2008) can not find the sign of the 
He II line in their spectrum during 88--132 days.

Here we investigate whether dust formation in the postshock gas around
SN 2006jc is possible or not, referring to the X-ray observations and 
the hydrodynamical calculations for the ejecta-CSM interaction.
Observations with {\it SWIFT} and {\it Chandra} satellites (Immler et 
al. 2008) have shown that the X-ray luminosity is in the range of 
10$^{39}$--10$^{40}$ ergs s$^{-1}$ from day 36 through day 180 with a 
gentle rise of the X-ray emission until 110 days.
Immler et al. (2008) have estimated the density of the postshock gas 
shell to be $\sim$$4 \times 10^{-17}$ g cm$^{-3}$, assuming that all of 
the measured X-rays at peak are emitted from the dense shocked shell.
However, the time evolution of the X-ray luminosity indicates that the 
medium around SN 2006jc has a quite smooth density profile. 
Tominaga et al. (2007) have performed the hydrodynamic calculations to 
reproduce the observed X-ray light curve and found that the density 
profile of the preshock CSM should be 
%\textcolor{red}{
$\rho_0 = 2.75 \times 10^{-19}$ g cm$^{-3}$ for 
$r < 2.2 \times 10^{16}$ cm and $\rho = \rho_0 (r/2.2 \times 
10^{16}$cm)$^{-6}$ g cm$^{-3}$ for 
$r \ge 2.2 \times 10^{16}$ cm.
%}
Their result has shown that the gas density swept up by the forward and
reverse shocks is $\la 10^{-17}$ g cm$^{-3}$ at 50 days (see Figure 10 
in Tominaga et al. 2007), which is too low for dust grains to nucleate.
Thus, on the basis of the hydrodynamical simulations, we conclude that 
dust grains are not likely to be formed in the postshock CS gas.

On the other hand, it should be addressed that there are some studies that 
advocate formation of dust in the dense CS shell produced by the LBV-like 
outburst before two years (Di Carlo et al. 2007; Mattila et al. 2008).
Mattila et al. (2008) have shown that the observed X-ray luminosity is 
produced by the CS shell with the preshock density of (2.5--4) $\times 
10^{-15}$ g cm$^{-3}$ in aid of their calculations of shock evolution and 
that the NIR light curve is reproduced by the IR light echo of the dust 
newly formed in the dense, cooling CS shell.
However, formation process of dust heavily depends on the cooling history 
of the gas, and whether dust formation can be realized in the rapidly 
cooling postshock gas should be fully explored in the future work.
In the following sections, we shall persue the possibility of dust 
formation in the expanding ejecta of SN 2006jc.

\section{Dust Formation in the Ejecta of SN 2006jc}

\subsection{Model of SN 2006jc and calculation of dust formation}

The time at which dust condenses in the SN ejecta and the size and number 
density of the newly formed dust are sensitive to the time evolution of 
the gas temperature and density (Kozasa et al. 1989; Todini \& Ferrara
2001; Nozawa et al. 2003), given the elemental composition.
In the calculation of dust formation, we apply the results of 
hydrodynamics and nucleosynthesis calculations for the SN Ib model well 
reproducing the observed bolometric light curve of SN 2006jc 
(Tominaga et al. 2007);
the ejecta mass is 4.9 $M_\odot$ with the explosion energy of 
$E_{\rm ex}/10^{51}$ergs $=E_{51}=10$ and the ejected $^{56}$Ni mass of
0.22 $M_\odot$, where $^{56}$Ni is uniformly distributed in mass within
the He core.
The pre-explosion mass of the progenitor is 6.9 $M_\odot$.
The elemental composition in the ejecta has an original onion-like
composition, except for $^{56}$Ni.
The time evolution of the gas temperature in the freely expanding ejecta 
was calculated by solving the radiative transfer and the energy 
equations, taking account of the energy deposition from $^{56}$Ni and 
$^{56}$Co. 

Figure 1 shows the structures of the gas temperature (Fig. 1{\it a}) and 
density (Fig. 1{\it b}) within the He core of the SN 2006jc model at 50 
days ({\it red solid lines}) and 200 days ({\it red dashed lines}) 
after the explosion.
For comparison, shown are those for the SN II model with the progenitor 
mass $M_{\rm pr} = 25$ $M_\odot$ and $E_{51}=1$ ({\it blue solid} and 
{\it dashed lines} at 50 and 200 days, respectively; Umeda \& Nomoto 
2002), which has a He core of 5.5 $M_\odot$ comparable to that of SN 
2006jc.
The gas temperature in SN 2006jc decreases much earlier, and the density 
at a given time is more than three orders of magnitude lower than that
in the SN II.
This is because the expansion velocity in the He core of SN 2006jc is 
considerably high, reflecting the absence of the hydrogen envelope and 
a ten times higher explosion energy than SNe II with 
$E_{51}=1$.
The gas temperature in SN 2006jc drops down to $\sim$2000 K at 50 days
and $\sim$1000 K at 200 days.
Thus, we can expect early formation of dust grains between 50 days and 
200 days, since a typical dust condensation temperature ranges from 
1000 K to 2000 K.

The dust formation calculation is performed by applying a theory of 
non-steady state nucleation and grain growth described in Nozawa et al. 
(2003).
In the calculations, we assume the sticking probability of $\alpha_s=1$,
and that the temperature of a small cluster consisting of up to a few 
tens atoms is the same as the gas temperature.\footnote{
There is a typographical error in the equation (3) for the nucleation 
rate in Nozawa et al. (2003); replace $c_{1j}$ with $c_{1j}^2$.}

\subsection{Results of dust formation calculations}

Figures 2{\it a} and 2{\it b} show the results of calculation.
Figure 2{\it a} indicates the condensation time of the newly formed dust 
defined as the time when the nucleation rate reaches the maximum (see 
Nozawa et al. 2003). 
It can be seen that in the outermost C-rich layer of the ejecta (see 
Fig. 3{\it c} in Tominaga et al. 2007), C grains condense at very early 
times of 40--60 days. 
This is due to the rapid decrease of the gas temperature in the ejecta
as well as the relatively high condensation temperature of C grains. 
It should be emphasized that the condensation times of C grains are in 
good agreement with the onset time of dust formation observed in SN 
2006jc.
Furthermore, the condensation of C grains can explain both the 
blueshift of He I lines and the entire fading of Ca II and O I lines 
because the formation site is in the region between the He-rich CSM
and the O-rich layer in the ejecta, though they cannot absorb the 
zero-velocity component of narrow He I emission lines efficiently.

After C grains are formed, Al$_2$O$_3$, Mg$_2$SiO$_4$, MgSiO$_3$,
SiO$_2$, Fe$_3$O$_4$, MgO, and FeO grains form in the O-rich layer at 
90--150 days in this order, and FeS and Si grains condense in the
Si-S-rich layer at $\sim$200 days.
Since $^{56}$Ni is uniformly mixed throughout the ejecta, Fe$_3$O$_4$
and FeO grains are produced instead of Fe grains;
Fe grains are expected to form in the innermost Fe-Ni core unless 
$^{56}$Ni intrudes into the outer layers (Nozawa et al. 2003).
However, the mixing of elements is likely to be at the knotty level
rather than at the microscopic atomic level, as suggested by the
observations of the Cas A supernova remnant (Douvion et al. 2001; 
Ennis et al. 2006; Rho et al. 2008).
If $^{56}$Ni is not mixed at the atomic level, the formation of Fe 
grains is expected as well.
Nevertheless, Fe grains, which have a high energy barrier for nucleation 
and a resulting low condensation temperature ($\la$800K), cannot form in 
the ejecta considered here because the gas density is too low at the
time when the gas cools down to 800K ($\sim$250 days).
Also, Fe atoms originally existing in the O-rich layer are not locked 
into grains significantly (condensation efficiency of $\la$10$^{-7}$) 
owing to their small abundances.
Considering these uncertainties, we rule out Fe$_3$O$_4$ and FeO grains 
as the newly formed dust species. 

As is expected from the low gas density as well as the rapid cooling of 
gas in the ejecta, the average radius of each dust species given in 
Figure 2{\it b} is considerably small, being less than 0.01 $\micron$ 
for all grain species. 
On the other hand, the condensation efficiency of dust is almost unity, 
and the amount of dust formed is relatively large. 
The total mass of newly formed dust is 1.45 $M_\odot$, and the mass of C 
grains (0.7 $M_\odot$) accounts for about a half of the total mass. 
The mass of each dust species $M_{1,j}$ obtained by the dust formation 
calculation is summarized in Table 1. 
Other major grain species in mass are SiO$_2$ (0.23 $M_\odot$), Si
(0.2 $M_\odot$), and MgSiO$_3$ (0.16 $M_\odot$). 

It should be noted here that the dust formation process as demonstrated
above is not influenced by the interaction between the ejecta and the
CSM.
Our results show that dust grains can not condense in the region of 0.005 
$M_\odot$ from the outer edge of the ejecta where the gas density is too 
low.
Although the ejecta-CSM interaction causes the reverse shock penetrating
into the ejecta, the hydrodynamic simulation shows that the reverse
shock sweeps up only $\sim$0.011 $M_\odot$ of the outermost region in the 
ejecta by 50 days and $\sim$0.053 $M_\odot$ by 200 days (see Fig. 10 in 
Tominaga et al. 2007).
Thus, the reverse shock 
%\textcolor{red}{
sweeps up 0.006 $M_\odot$ 
from the outer edge of the dust formation region at day 50, and 0.048 
$M_\odot$ at day 200.  
The mass of C grains within the region swept up by 200 days is only 
$1.3 \times 10^{-3}$ $M_\odot$, which is much smaller than 0.7 $M_\odot$ 
at the time of formation, thus almost all of C grains can form in the 
freely expanding ejecta.
%\textcolor{red}{
Since the radii of C grains formed in the outermost region are smaller 
than 10 \AA~(see Fig. 2b), the timescale of dust destruction by sputtering 
is estimated to be $\la$6 hours (Nozawa et al. 2006) in the shocked gas 
with density of $\sim$$10^{-18}$ g cm$^{-3}$ (Tominaga et al. 2007).
Hence, thermal emission from shock-heated grains can be negligible.
%} 

%Although the ejecta-CSM interaction causes the reverse shock penetrating
%into the ejecta, the hydrodynamic simulation shows that the reverse
%shock sweeps up only $\sim$0.003 $M_\odot$ of the outermost region in the 
%ejecta by 50 days and $\sim$0.014 $M_\odot$ by 200 days (see Fig. 10 in 
%Tominaga et al. 2007).
%Thus, the reverse shock cannot reach the outermost region of dust 
%formation before 50 days, and C grains can form in the freely expanding 
%ejecta.
%At 200 days, C grains within the region of 0.008 $M_\odot$ from the outer 
%edge of dust formation region are processed and heated by the collision 
%with the reverse shock, but the mass of C grains included in this region 
%is only $1.4 \times 10^{-6}$ $M_\odot$, which has no effect on the 
%following discussion.

\section{Evaluation of Dust Temperature and Mass in the Ejecta}

The NIR to mid-IR (MIR) observations with {\it AKARI} (Sakon et al. 
2007) and {\it MAGNUM} (Minezaki et al. 2007) have detected thermal 
emission from dust in SN 2006jc at $\sim$200 days.
Sakon et al. (2007) have concluded that the obtained NIR spectrum is
fitted by amorphous carbon grains of 800 K with a mass of $6.9 \times 
10^{-5}$ $M_\odot$, assuming optically thin thermal radiation.
They have also shown that the MIR excess emission over the amorphous 
carbon grains of 800 K can be explained either by silicate and silica 
grains of 700--800 K with $\sim$$10^{-4}$ $M_\odot$ or amorphous carbon 
grains of 320 K with $2.7 \times 10^{-3}$ $M_\odot$.
In this section, referring to the grains species calculated in \S~3 as 
possible condensates in the ejecta and deriving the dust temperature,
we evaluate the dust mass by reproducing the spectral energy 
distribution observed at day 200 with {\it AKARI} and {\it MAGNUM}.

\subsection{Temperature of dust}

The equilibrium temperature of dust in the ejecta is determined by the 
balance between heating and cooling through radiation and collision 
with gas.
In order to calculate the heating due to the absorption of ambient 
radiation, we evaluate the flux $F_{\lambda}(r,t)$ at a time $t$ and a 
position $r$ within the ejecta, assuming that the total flux $F(r,t)$ 
calculated by Tominaga et al (2007) radiates as a blackbody with
temperature $T_{\rm BB}(r,t)$.

Given the gas temperature $T_{\rm gas}(r,t)$ and number density $n_{\rm
gas}(r,t)$ in the ejecta, the equilibrium temperature of dust 
$T_{\rm d}(r,t)$ at a given time and at a given position is determined 
by implicitly solving the equation 
\begin{eqnarray}
4 \pi a^2 \sigma_{\rm B} T_{\rm d}(r,t)^4 \langle 
Q_\lambda(a, T_{\rm d}) \rangle 
=\frac{F(r,t)}{\sigma_{\rm B} T_{\rm BB}^4} \int_0^{\infty} 
\pi a^2 Q_\lambda(a) B_\lambda(T_{\rm BB}) d\lambda \nonumber \\
+ 4 \pi a^2 n_{\rm gas} 
\left(c_{\rm v} + \frac{1}{2} k \right)
\left[
T_{\rm gas} \left(\frac{k T_{\rm gas}}{2 \pi \mu m_{\rm H}} 
\right)^{\frac{1}{2}} - T_{\rm d} \left(\frac{k T_{\rm d}}{2 \pi 
\mu m_{\rm H}} \right)^{\frac{1}{2}} \right],
\end{eqnarray}
where $a$ is the radius of dust, $\sigma_{\rm B}$ is the 
Stefan-Boltzmann constant, $\langle Q_\lambda (a, T) \rangle$ is the 
Planck-mean of absorption coefficient, $B_\lambda(T)$ is the Plank 
function, $k$ is the Boltzmann constant, 
$c_{\rm v}$ is the heat capacity of a gas molecule 
($c_{\rm v} = 3k/2$ for a monatomic molecule), $\mu$ is the mean 
molecular weight, and $m_{\rm H}$ is the atomic mass unit.
Here we consider that the gas colliding with dust is completely 
accommodated and is ejected from the surface with the same temperature 
as the dust (see Landau \& Pitaevski 1981).
In the calculations we assume $T_{\rm BB}=5000$ K throughout the
ejecta, regardless of time, and $a=0.01$ $\mu$m. 
Since in the ejecta considered here, the total flux at day 200 is 
$F \sim 10^6$ erg s$^{-1}$ cm$^{-2}$ almost uniformly over the He core,
the energy transfered through radiative process is much larger than 
that by collision with gas, and thus we can neglect the collisional 
heating and cooling. 
Note that in this case the calculated dust temperature is independent
of dust size because $Q_\lambda(a)/a$ does not depend on the radius for 
$a \la 0.05$ $\mu$m.
The references for the optical constants of each dust species 
used in the calculations are given in Table 1.

We also note that the effect of stochastic heating can be negligible as 
long as we assume a blackbody radiation with $T_{\rm BB} \sim 5000$ K 
under the given total flux.
We have confirm that for C grains of 10 (20) \AA, stochastic heating 
only makes the dust temperature distribute symmetrically around the 
equilibrium temperature with the temperature width of $\sim$$\pm$100 (30) 
K.
Accordingly, the resulting emissivity of 10 \AA-sized C grains does not 
change significantly at wavelengths longer than 2 $\mu$m, compared with 
that derived for equilibrium temperature; its difference is 10 \% at 2 
$\mu$m and is less than 2.5 \% at $\ge$ 3 $\mu$m.
Therefore, stochastic heating does not affect the results of the IR 
spectrum energy distribution calculated in \S~4.2.

Figure 3 shows the temperatures of all dust species at 200 days 
({\it solid lines}), along with the temperature of C grains at 60 days
({\it dashed lines}).\footnote{Note that in the calculation we exclude 
Si grains; 
the temperature of Si grains at the condensation time of $\sim$200 days
is about 300 K higher than the gas temperature.
When we consider the non-LTE effect for dust formation, its condensation 
time is delayed to 240 days, as discussed in \S~5.1, which is later than 
the observation date of {\it AKARI} and {\it MAGNUM}.}
The temperature of the C grains is 1200--1700 K at 60 days and is higher
for those formed in the inner region.
Note that this high temperature is in good agreement with the
temperature of dust responsible for the red/NIR continuum appearing in
SN 2006jc two months after the explosion (Smith et al. 2008).
At 200 days, the C grains cool down to 500--600 K, which is also 
considered to be correspondent to the temperature of 800 K for amorphous 
carbon grains suggested by the observations with {\it AKARI} and 
{\it MAGNUM} (Sakon et al. 2007), although the calculated temperature is 
relatively low.
Thus, the calculated temperature of C grains can reasonably explain the
time evolution of the dust temperature indicated by the observations 
toward SN 2006jc from day 50 through day 200.

On the other hand, silicate grains have a temperature of only 100--200 K 
at day 200, reflecting much lower absorption efficiency at optical 
wavelengths than at IR wavelengths.
This low temperature cannot explain the MIR excess emission over the 
amorphous carbon grains with 800 K observed by {\it AKARI} (Sakon et
al. 2007), and we can conclude that neither silicate nor silica 
grains newly formed in the ejecta can be the carriers of the MIR 
excess emission. 
In addition, our result shows that the temperature of C grains spans the 
range from 500 to 600 K.
Therefore, if the MIR excess emission would be attributed to the 
amorphous carbon grains with 320 K, they might not be the grains formed 
in the ejecta but could be the pre-existing CS grains heated by the SN 
outburst, as suggested by Sakon et al. (2007).

In the calculations we assumed the ambient radiation field as a
blackbody with $T_{\rm BB}=5000$ K. 
The calculated dust temperatures at day 200 is not sensitive to 
$T_{\rm BB}$ as long as $T_{\rm BB}$ is in the range of 5000 K to 6000 
K;
for $T_{\rm BB} = 6000$ K, the temperature of MgSiO$_3$ and 
Mg$_2$SiO$_4$ grains increases by $\sim$50 K compared with that for 
$T_{\rm BB} = 5000$ K, while the increase of temperature of the other 
grains species is at most 30 K.
Here, it should be kept in mind that, being different from normal SNe
Ib, SN 2006jc shows the strange optical spectrum in the early times
until day 120, with a bright blue continuum whose origin is not clear 
(Foley et al. 2007, Pastorello et al. 2007).
This indicates that the spectrum of radiation at the early epoch within 
the ejecta of SN 2006jc is not also expected to be approximated by a 
blackbody radiation. 
Nevertheless, as shown above, we can reasonably explain the time
evolution of temperature of C grains indicated by the observations.
Thus, we adopt the results for $T_{\rm BB} = 5000$ in the following 
discussions.

\subsection{IR spectral energy distribution and dust mass}

In this subsection, we calculate thermal emission from dust grains based 
on the dust temperatures presented in Figure 3, and estimate the dust 
mass necessary for reproducing the spectral energy distribution (SED)
at day 200 observed with {\it AKARI} and {\it MAGNUM}.

Taking into account the self-absorption of dust, the observed flux
density $F_{\lambda}^{\rm obs}$ of thermal emission from dust grains
formed in the ejecta is evaluated by
\begin{eqnarray}
F_{\lambda}^{\rm obs} = \frac{1}{4 \pi D^2} \int^R_0 \sum_j 
4 \pi r^2 m_{{\rm d}, j}(r) \kappa_{\lambda, j} B_\lambda(T_{\rm d}(r)) 
\exp[-(\tau_\lambda (R)-\tau_\lambda(r))] dr,
\end{eqnarray}
where $D = 25.8$ pc is the distance from observers (Pastorello et
al. 2007), $R$ is the outermost radius of the dust-forming region in the
ejecta, $m_{{\rm d},j}(r)$ is the mass of dust species $j$ per volume at 
the position $r$, and $\kappa_{\lambda,j} = 3 Q_{\lambda,j}(a)/4 a
\rho_j$ is the mass absorption coefficient, and the dust bulk density 
$\rho_j$ is taken from Nozawa et al. (2006). 
Note that $\kappa_{\lambda,j}$ does not depend on the size as long as 
$a \la 0.05$ $\mu$m.
The optical depth $\tau_\lambda$ is given by
\begin{eqnarray}
\tau_\lambda(r) = \int^r_0 \sum_j m_{{\rm d}, j}(r) \kappa_{\lambda, j}
 dr.
\end{eqnarray}

The magenta solid line in Figure 4 is the SED calculated by adopting 
the mass of each dust species $M_{1,j}$ given in \S~3.2, and the SED is 
dominated by thermal emission from C grains existing in the outermost 
ejecta as the result of the large optical depth due to their large mass.
It is clear that the derived IR spectrum largely exceeds the 
observational data at $\ga$3 $\mu$m and cannot reproduce the observed
SED.

The NIR to MIR spectrum can be fitted using the derived grain 
temperatures because dust temperature does not depend on the mass of 
dust as can be seen from Equation (1), so the dust mass can be used as 
a free parameter in order to reproduce the observations.
We calculate thermal emission 
from dust grains by taking the mass of each grain species at the 
position $r$ as $f_j m_{{\rm d},j}(r)$ and fit the SED to the 
observational data, where the value of a free parameter $f_j (\le 1)$ is 
assumed to be constant for a given dust species.
The best fitted result is displayed by the red solid line in Figure 4 
and can well explain the observed NIR to MIR spectrum.
The contribution of each grains species is depicted by the dashed lines,
for which we adopt the masses of C and FeS grains and the upper mass 
limits of the other dust species given as $M_{2,j}$ in Table 1.
Here the mass of C grains is $5.6 \times 10^{-4}$ $M_\odot$, reduced by 
a factor of 1250 in comparison with the mass obtained by the dust
formation calculation.
FeS grains can also contribute to the NIR spectrum because of its high 
temperature ($\sim$900 K), and its mass is reduced to $2 \times 10^{-3}$ 
$M_\odot$ ($f_j = 0.03$).
On the other hand, silicates and oxide grains do not contribute to the 
NIR spectrum because their temperatures are too low (100--200 K).
Furthermore, the large optical depth of MgSiO$_3$ and Mg$_2$SiO$_4$ at 
MIR wavelengths prevents us from constraining the mass of these grain 
species.
Thus, we can not defy the existence of a large amount of cold dust 
reaching up to $\sim$0.5 $M_\odot$ in the ejecta from the spectral 
fitting, and the dust mass derived under the optically thin assumption 
should be considered to be the lower limit (see also Meikle et al. 
2007).

We also investigate the effect of the presence of amorphous carbon 
grains in the CSM on the observed spectrum, which have been suggested 
to contribute to the MIR spectrum at $\ga$200 days (Sakon et al. 2007; 
Mattila et al. 2008).
In the calculations, we take the temperature of CS carbon grains to be 
320 K (Sakon et al. 2007).
If their mass is smaller than $2.7 \times 10^{-4}$ $M_\odot$, the 
best-fitted spectrum in Figure 4 is not influenced by the pre-existing 
C grains.
Further increase of CS dust mass can also reasonably reproduce the 
observed IR spectrum as seen from Figure 5, where the mass of newly 
formed and pre-existing C dust is $2.24 \times 10^{-4}$ $M_\odot$ and 
$1.4 \times 10^{-3}$ $M_\odot$, respectively (the mass of other newly 
formed dust is the same as that in Figure 4).
Therefore, the spectral fitting can allow the presence of low-temperature 
C grains in the CSM.

The emission from CS grains can be the IR light-echo but not the thermal 
radiation from dust heated by the collision with the forward shock. 
This is readily demonstrated by considering the radius of dust cavity
created by the evaporation of pre-existing CS grains due to strong 
radiation from a SN.
Taking the evaporation temperature of C grains to be 1800 K and using 
the SN peak luminosity of $10^{42.7}$ ergs s$^{-1}$ with the blackbody 
temperature of $\sim$6000 K (Tominaga et al. 2007), we estimate the 
radius of dust cavity to be $\simeq3 \times 10^{16}$ cm.
Because the estimated radius of dust cavity is comparable with the 
radius of the forward shock at $\sim$200 days, the forward shock could 
have not swept a significant amount of CS dust.
If we consider the radiation from the shock breakout as the heating 
source, which is expected to be brighter and have a higher blackbody 
temperature, the dust cavity may be extended so far that any CS dust 
can not be swept up by the forward shock at $\sim$200 days.
Note that in this spectral fitting, the mass of newly formed C grains 
depends on the temperature and mass of CS dust, and in the following 
discussion, we refer to the mass of C grains derived by the spectral 
fitting without CS dust in Figure 4.

\section{Discussion}

\subsection{Mass of dust formed in the ejecta}

As shown in \S~4.2, the mass of C grains (0.7 $M_\odot$) obtained by the 
dust formation calculation is three orders of magnitude larger than that 
($5.6 \times 10^{-4}$ $M_\odot$) estimated from the spectral fitting to 
the observations.
One of the reasons for this difference is considered to be the clumpy 
structure in the ejecta.
In order to evaluate the dust mass necessary for reproducing the 
observed spectrum of SN 2003gd, Sugerman et al. (2006) performed the 
radiative transfer calculation taking account of the clumpy structure 
and found an increase in the estimated mass of dust in the ejecta by 
about one order of magnitude, compared with the mass derived for the 
uniform distribution of dust.
Nevertheless, the consideration of dust clumping does not seem to be 
sufficient to overcome the large gap.
Thus, the difference in dust mass should stem from the process of dust 
formation in the ejecta.

One of the processes influencing on dust formation is the non-local
thermal equilibrium (non-LTE) effect. 
As seen from Figures 1 and 3, the dust temperature is not the same as 
the gas temperature; at 200 days the temperature of C grains (500--600 
K) and silicate and oxide grains (100--200 K) is lower than the gas 
temperature with 700--1000 K.
In the ejecta with strong radiation field, the temperature of small 
clusters is expected to be higher than the gas temperature depending 
on the optical property. 
In this case, the non-LTE effect retards or hinders the formation of 
dust grains.
Assuming that the optical properties of small carbon clusters are the
same as those of small C grains, and applying a formula of nucleation
rate taking into account the non-LTE effect by Kozasa et al (1996), 
we realized that the non-LTE effect does not significantly affect the 
formation of C grains; the non-LTE effect only retards the formation of 
C grains by less than ten days and does not affect the mass of C grains 
because the temperature of small carbon clusters is only 50 K higher 
than that of gas. 
Meanwhile, the non-LTE effect delays the formation of Si grains from 
$\sim$200 days to $\sim$240 days. 
This is the reason why we exclude Si grains in reproducing the observed 
SED at day 200 in \S~4.2.

Another process affecting the formation of dust in the ejecta is the
energy deposition on small-sized clusters through the latent heat 
deposition at the condensation as well as the collisions with high 
energy photons and electrons.  
In the ejecta rich in condensible elements, the destruction of
small-sized clusters by the deposition of latent heat could be 
significant unless the deposited energy is released by collisional 
and/or radiative processes efficiently. 
Also, at an early epoch after the explosion, energetic photons and
electrons generated from the decays of radioactive elements $^{56}$Ni 
and $^{56}$Co prevail abundantly throughout the ejecta. 
Since the nucleation rate is determined by the balance between 
growth and destruction rate of small-sized clusters, 
the energy deposition makes small-sized clusters unstable against 
the growth and as a result reduce the nucleation rate. 

Although the nucleation rate can be reduced by increasing the critical
cluster size as demonstrated by Bianchi \& Schneider (2007), no reason 
is there to cut the number of seed nuclei by increasing the critical
size arbitrarily, apart from the uncertainty inherent in the standard 
nucleation theory arising from the evaluation of chemical potential of 
small-sized clusters. 
Thus, we examine the effect of energy deposition for dust formation at 
early times observed in SN 2006jc by reducing the sticking probability 
$\alpha_s$ for simplicity and comparing with the mass of dust evaluated 
in \S~4.2.
If we take $\alpha_{\rm s} \sim 3 \times 10^{-3}$, we can obtain the 
mass of C grains of $\sim$$5 \times 10^{-4}$ $M_\odot$, while 
$\alpha_{\rm s}$ is $\sim$$0.3$ to produce FeS grains of $2 \times
10^{-3}$ $M_\odot$.
In this case, the condensation times of C and FeS grains is 40--67 
days and 190--200 days, respectively.
While the destruction rate of small-sized clusters depends on the 
properties of materials forming dust grains, the difference in the 
sticking probability may be caused by the collisions with energetic
photons and electrons, reflecting the difference in the condensation
time of C and FeS grains; energetic photons and electrons are more 
abundant at day 50 than at day 200.
The details of these microscopic processes are beyond the scope of 
this paper and will be investigated in the future study on the dust
formation in SNe. 

\subsection{Early condensation of dust in SNe Ib}

In \S~3.2, we show that the early formation of dust around 50 days is 
possible in SN 2006jc whose progenitor has lost most of He envelope 
(Foley et al. 2007; Tominaga et al. 2007). 
However, other dust-forming SNe Ib had not show such an early formation 
of dust; 
SN 1990I displayed the diagnostics of dust formation around 230 days 
(Elmhamdi et al. 2004), though the condensation time is relatively
earlier than that of SN II.
A peculiar Type Ib SN 2005bf shows a rapid decline of the light 
curve from 50 days after the explosion, but did not exhibit the NIR
brightening in contrast to the case of SN 2006jc (Tominaga et al. 2005).
Maeda et al. (2007) have shown that the quickly faded light curve of SN 
2005bf can be explained by the newly born magnetized neutron star at 
the center. 
On the other hand, they have demonstrated that the blueshifted profiles 
of O I and Ca II observed at $\sim$270 days can be explained by the
obscuration of dust in the ejecta. 
If the blueshift of emission lines in SN 2005bf is attributed to the 
newly formed dust grains, the time of dust formation is considered to be 
later than 100 days and not to be so early as in SN 2006jc, although the 
epoch of dust formation in SN 2005bf was not specified by the
observations.

The results of calculations revealed that the early formation of dust in 
SN 2006jc can be realized by the rapid decline of gas temperature,
reflecting the large ratio of the explosion energy ($E_{51}= $10) to the
ejected mass ($M_{\rm ej}$ = 4.9 $M_\odot$);
the ratio $E_{51}/(M_{\rm ej}/M_\odot) \simeq 2$ is much greater than
0.125--0.25 for SN 2005bf (Tominaga et al. 2005) and $\simeq$0.3 for SN 
1990I (Elmhamdi et al. 2004).
Unfortunately, no literature has reported the explosion energy of Type 
IIn SN 1998S, which had a time of dust condensation of $\sim$230 days 
(Gerardy 2000). 
Using the ejecta mass estimated from the He core of 4--6 $M_\odot$ 
(Fassia et al. 2001; Pooley et al. 2002) and hydrogen envelope of 0.1--2 
$M_\odot$ (Fassia et al. 2000; Anupama et al. 2001) and taking
$E_{51}=1$ as a typical explosion energy, we can, however, deduce the
ratio to be $E_{51}/(M_{\rm ej}/M_\odot) =$ 0.125--0.25 for SN 1998S,
which is similar to or slightly less than those for SN 2005bf and SN 
1990I.
On the other hand, the ratio for the dust forming SNe II is of order
0.1: 0.1--0.2 for SN 1987A (Shigeyama et al. 1988; Woosley 1988), 
0.05--0.1 for SN 1999em (Elmhamdi et al. 2003), and $\simeq$0.12 for 
SN 2003gd (Hendry et al. 2005).
Dust formation has not been yet observed on SNe Ic. 
However, the early dust formation in SNe Ic is also expected because SNe Ic 
may well have $M_{\rm ej}$ similar to SNe Ib.

In conclusion, we can speculate that as the envelope of star at explosion 
is smaller and the explosion energy is higher, the condensation time of 
dust tends to be earlier.
In order to prove these hypothesises and to investigate formation process 
of dust in SNe, we need further observational and theoretical studies on 
dust formation in various types of SNe.
In particular, the UV to MIR observations at early times for SNe Ib/c are 
promising to clarify the formation process of dust because its 
condensation time is much earlier than that for SNe II, as is shown above.

\section{Summary}

We investigate formation of dust grains in the ejecta of SN 2006jc by 
applying the SN model that can nicely reproduce its light curve.
We find that formation of dust in SN 2006jc is possible at early times, 
thanks to the rapid decrease of the gas temperature in the ejecta.
The condensation of C grains is realized in the C-rich layer at very 
early times of $\sim$50 days after the explosion.
This is in good agreement with the time of the onset of dust formation
indicated by the observations.
In addition to C grains, the other grain species such as silicate and 
oxide grains are formed until $\sim$200 day, with the average radii less 
than 0.01 $\micron$.
We suppose that formation of dust earlier than SNe II is common to the 
hydrogen-deficient SNe and that SN 2006jc is its extreme case.

Furthermore, we calculate the dust temperature by considering the
radiation field in the ejecta at the early epoch as a blackbody with 
$T_{\rm BB} = 5000$ K.
C grains have a temperature of $\sim$1600 K at 60 days and cools down to 
$\sim$600 K at 200 days. 
This temperature of C grains can reasonably explain the time evolution
of dust temperature suggested by the observations toward SN 2006jc
from day 50 through day 200.
The temperatures of silicates and oxide grains are 100--200 K at day 200
and are too low to contribute to the MIR spectrum observed with 
{\it AKARI}.

The comparison of the calculated SED with the observations of {\it
AKARI} and {\it MAGNUM} shows that dust grains contributing to NIR/MIR
flux are C and FeS grains whose masses are $5.6 \times 10^{-4}$ 
$M_{\odot}$ and $2 \times 10^{-3}$ $M_{\odot}$, respectively.
The masses of hot C and FeS grains obtained by the dust formation 
calculation are considerably larger than that estimated from the 
observations.
This difference in dust mass may be explained by the destruction of 
small-sized clusters by the collisions with energetic photons and
electrons.
On the other hand, 
we can not constrain the mass of low-temperature silicate and oxide 
grains in the ejecta from the spectral fitting.
The spectrum fitting also shows that the observed MIR spectrum can be 
attributed to pre-existing CS carbon grains with the mass up to 
$\sim$$10^{-3}$ $M_\odot$.
In order to reveal the process of dust formation in the SN ejecta, it is
essential to carry out UV to IR observations of various types of SNe. 
In particular, SNe Ib/c are the objects suitable for detecting
signatures of dust formation by the optical to NIR observations because 
the condensation time of dust is much earlier than SNe II. 

\acknowledgments

The authors are grateful to the anonymous referee for critical comments
that are useful for improving the manuscript.
The authors thank H. Hirashita, T. T. Takeuchi, and
A. K. Inoue for their useful comments.
%\textcolor{red}{
This research has been supported in part by World Premier International 
Research Center Initiative (WPI Initiative), MEXT, Japan, and by the 
Grant-in-Aid for Scientific Research of the Japan Society for the 
Promotion of Science (18104003, 18540231, 19740094, 20340038, 20540226, 
20540227) and MEXT (19047004, 20040004).
These grants support Nozawa, Kozasa, Tominaga, Tanaka, Suzuki, Nomoto, 
Maeda, Umeda and Limongi's visit.
%}
%T.N. has been supported in part by a Grant-in-Aid for Scientific
%Research from the Japan Society for the Promotion of Sciences
%(19740094, 18104003).

\clearpage

\begin{figure}
\epsscale{0.7}
\plotone{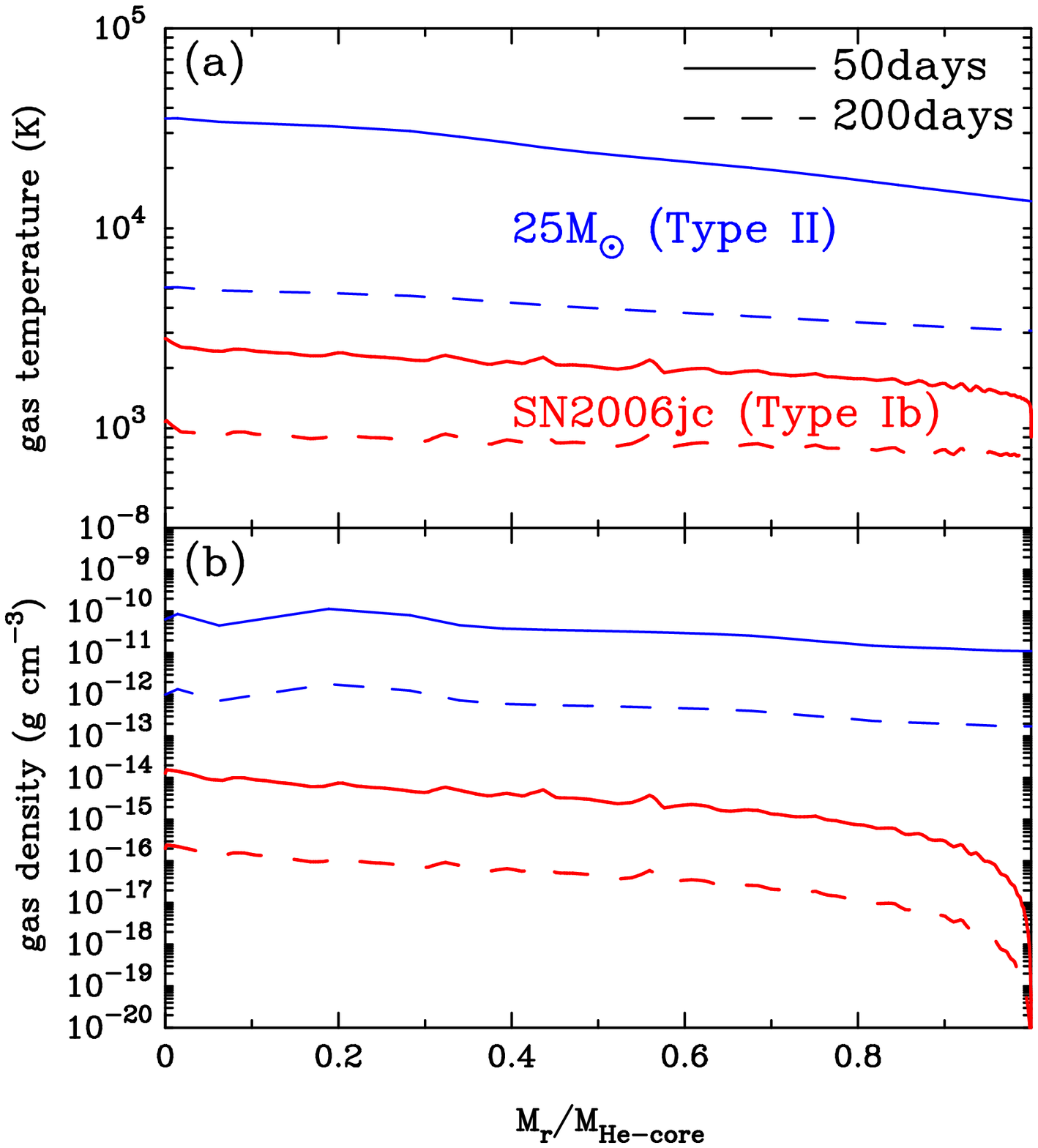}
\caption{
 Structures of (a) temperature and (b) density of the gas within the 
 He core of the SN 2006jc model ({\it red}) at day 50 ({\it solid
 lines}) and  day 200 ({\it dashed lines}) after the explosion.
 For comparison, shown are those for the SN II model ({\it blue}) with 
 $M_{\rm pr} = 25$ $M_\odot$ and $E_{51} = 1$ (Umeda \& Nomoto 2002), 
 which has a He core of 5.5 $M_\odot$ comparable to that of SN 2006jc.
 The mass coordinate is normalized by the mass of the He core.
\label{fig1}}
\end{figure}

\clearpage

\begin{figure}
\epsscale{0.7}
\plotone{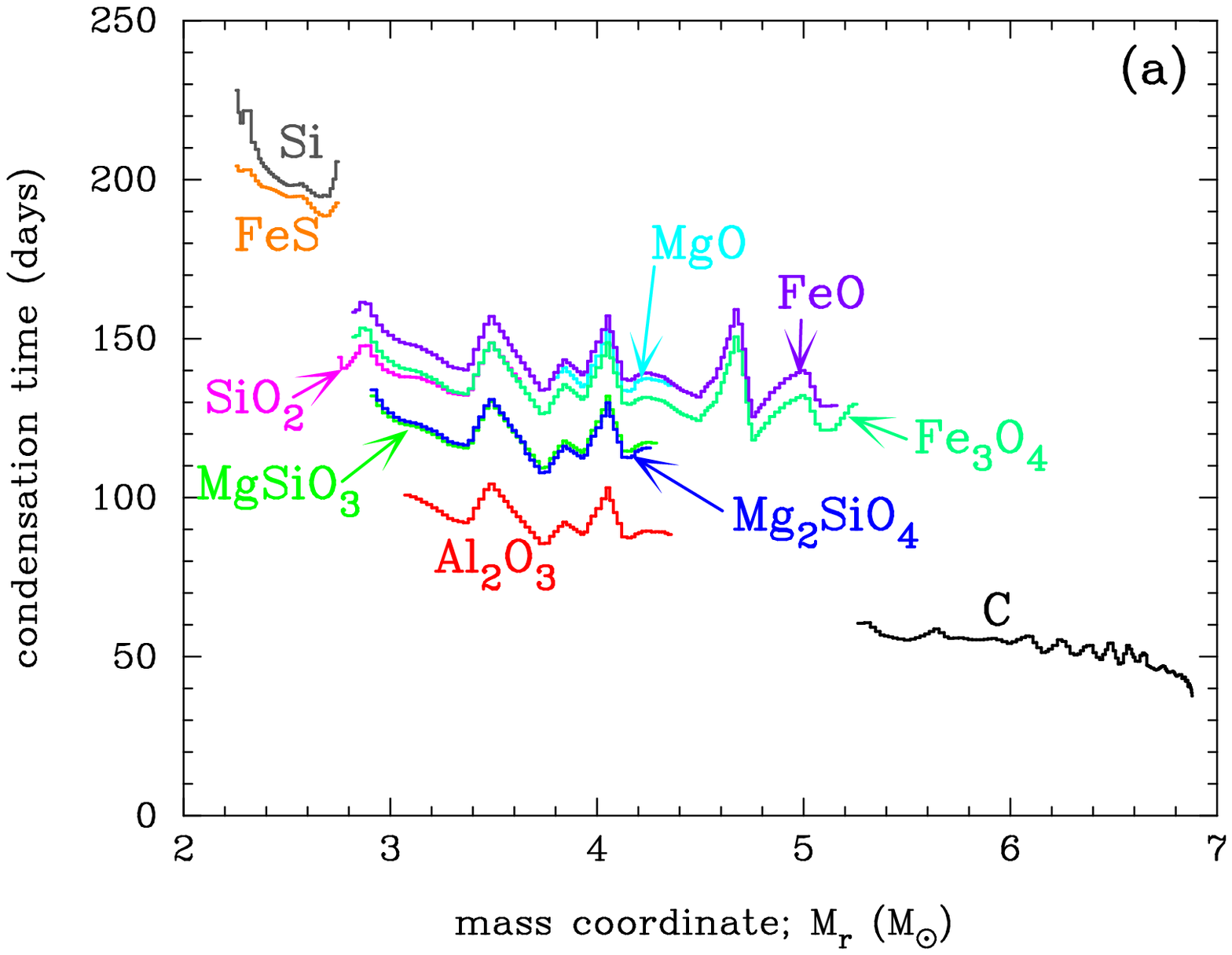}
\plotone{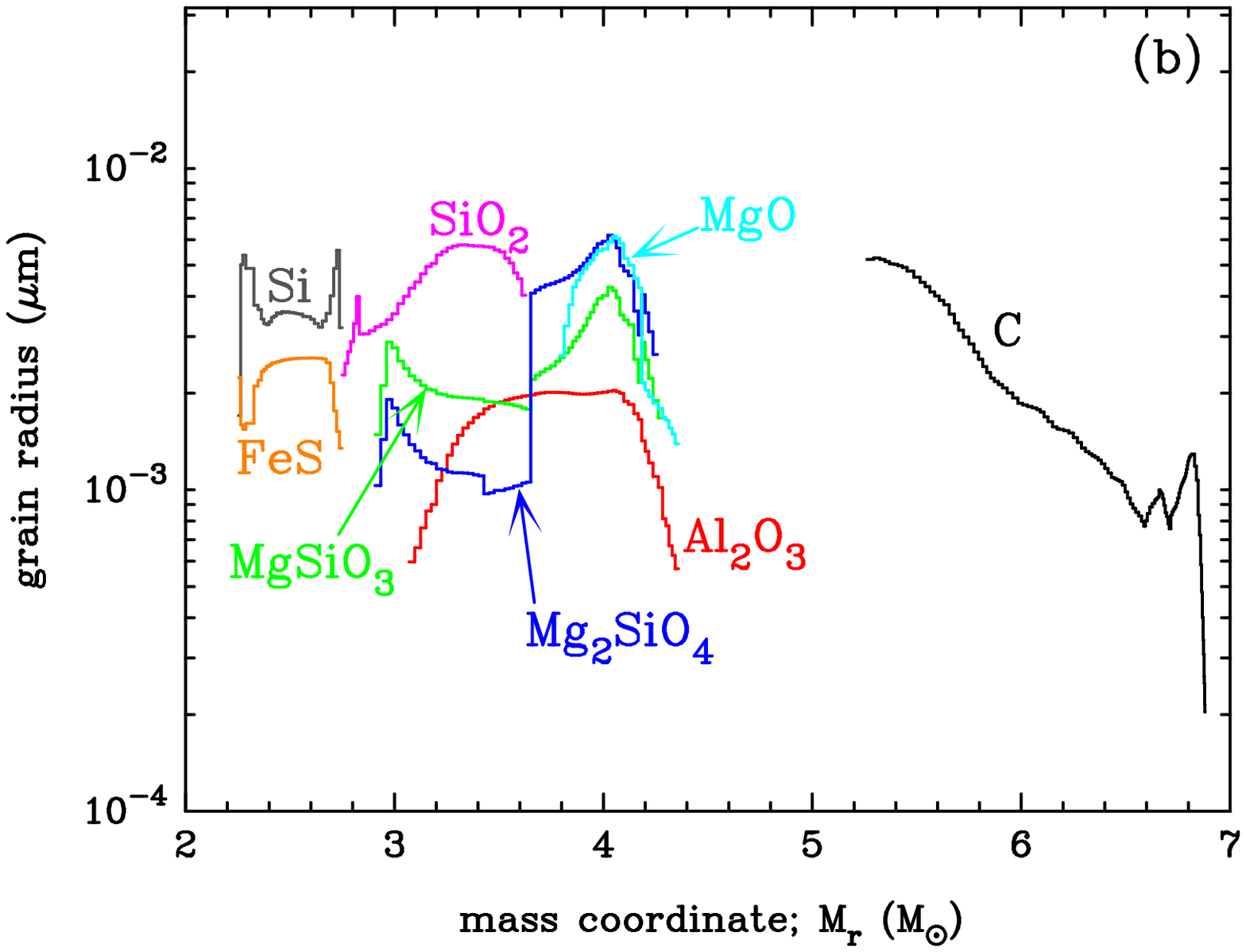}
\caption{
 ({\it a}) Condensation times and ({\it b}) average radii of dust grains 
 formed in the ejecta of SN 2006jc as a function of mass coordinate.
\label{fig2}}
\end{figure}

\clearpage

\begin{figure}
\epsscale{0.8}
\plotone{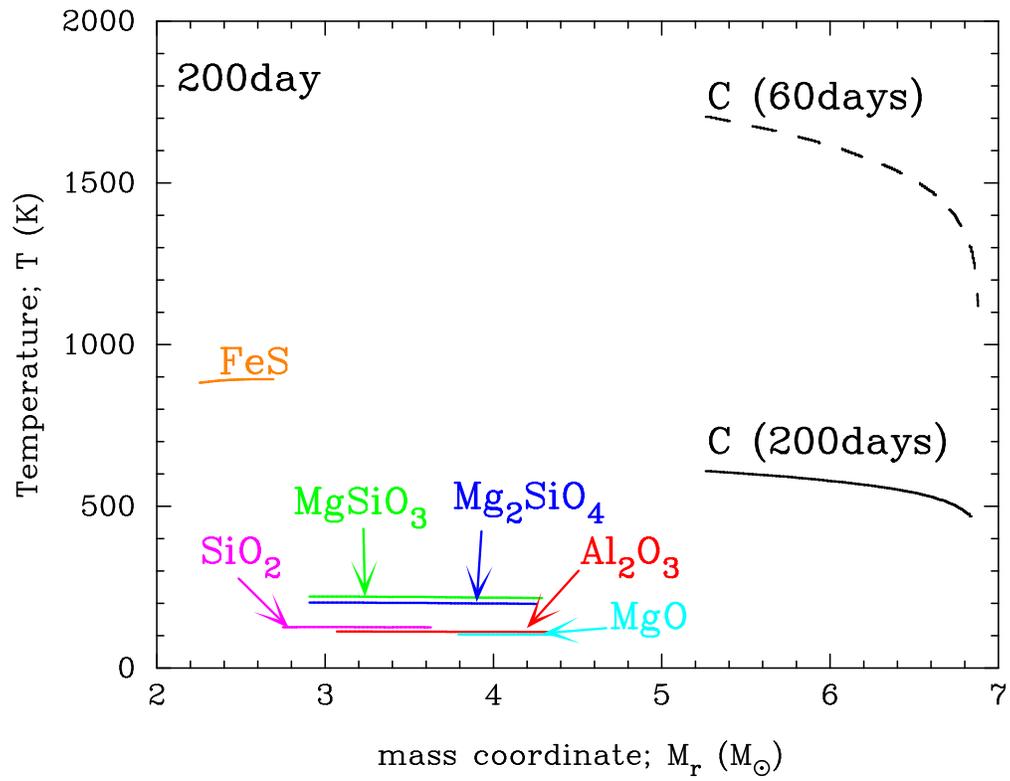}
\caption{Temperature of each dust species formed in the ejecta of SN
 2006jc at 200 days after the explosion.
 The dashed line depicts the temperature of C grains at 60 days.
\label{fig3}}
\end{figure}

\clearpage

\begin{figure}
\epsscale{0.8}
\plotone{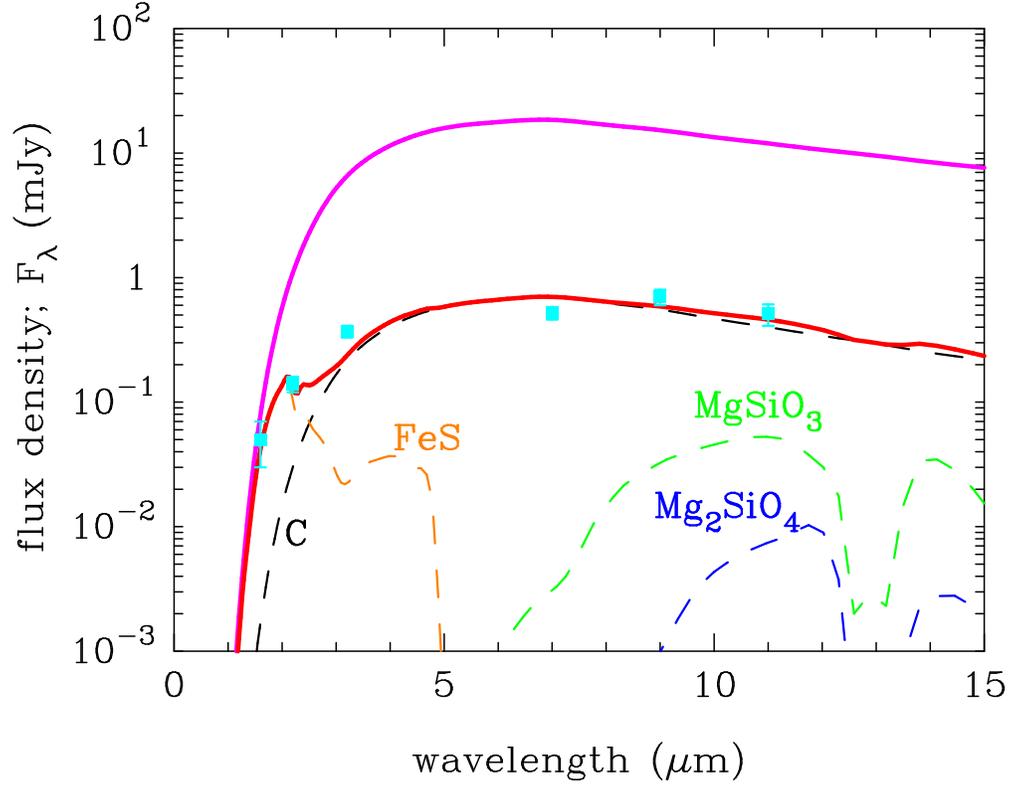}
\caption{ 
 Spectral energy distribution by thermal radiation from the newly formed
 dust.
 The magenta line is the spectrum obtained by adopting the mass of dust
 $M_{1,j}$ from the dust formation calculation. 
 The red line is the best fitted spectrum obtained by taking the dust 
 masses and upper mass limits $M_{2,j}$ given as in Table 1, for which 
 the contribution from each dust species is depicted by the dashed lines.
 The cyan symbols are the photometric data at $\sim$200 days by 
 {\it AKARI} (Sakon et al. 2007) and {\it MAGNUM} (Minezaki et al. 2007).
\label{fig4}}
\end{figure}

\clearpage

\begin{figure}
\epsscale{0.8}
\plotone{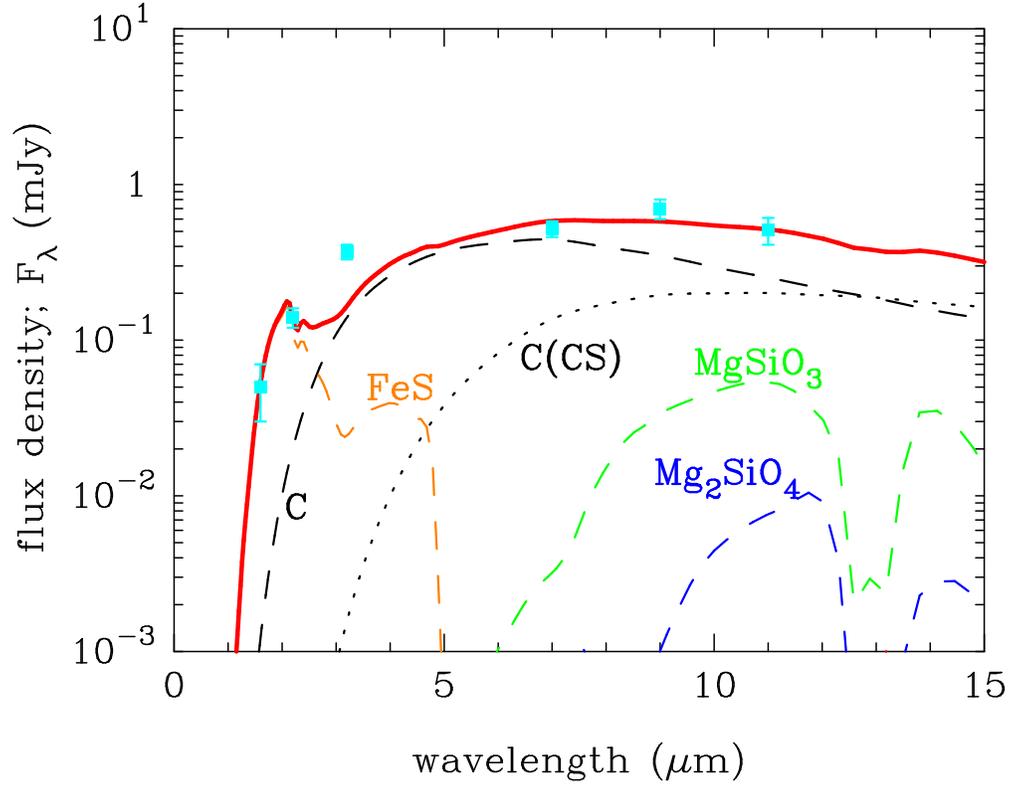}
\caption{ 
 Spectral energy distribution derived in the presence of CS carbon grains 
 ({\it thick red line}), where the mass of newly formed and pre-existing 
 C dust is $2.24 \times 10^{-4}$ $M_\odot$ ({\it black dashed line}) and 
 $1.4 \times 10^{-3}$ $M_\odot$ ({\it black dotted line}), 
 respectively. 
 The mass of other newly formed dust is the same as that in Figure 4.
 The cyan symbols are the photometric data at $\sim$200 days by 
 {\it AKARI} (Sakon et al. 2007) and {\it MAGNUM} (Minezaki et al. 2007).
\label{fig4}}
\end{figure}

\clearpage

\begin{deluxetable}{lccc}
\tablewidth{0pt}
\tablecaption{Mass of Each Dust Species and References for Optical Constants}
\tablehead{ 
\colhead{dust species} & \colhead{$M_{1, j}$ ($M_\odot$)} &
\colhead{$M_{2, j}$ ($M_\odot$)} & references 
}
\startdata
C             & 0.701 & $5.6 \times 10^{-4}$ & 1   \\
%C             & 0.701 \textcolor{red}{[0.693]} & $5.6 \times 10^{-4}$ & 1   \\
Al$_2$O$_3$   & 0.008 & $\le$0.008 & 2   \\
MgSiO$_3$     & 0.157 & $\le$0.157 & 3   \\
Mg$_2$SiO$_4$ & 0.082 & $\le$0.082 & 4   \\
SiO$_2$       & 0.229 & $\le$0.229 & 5   \\
MgO           & 0.010 & $\le$0.010 & 6   \\
%Fe$_3$O$_4$   & 0.021 & $\le$0.021 &     \\
%FeO           & 0.002 & $\le$0.002 &     \\
FeS           & 0.067 &      0.002  & 4   \\
Si            & 0.196 & --------    & 7   \\
Total         & 1.450 & $\le$0.489 &     
\enddata
\tablecomments{
The mass of grain species $j$ obtained by the dust formation calculation
is denoted by $M_{1,j}$. 
%\textcolor{red}{
%For C grains, the dust mass  at 200 days is also shown in square 
%brackets, since a small part of them are destroyed by the reverse shock,
%}
The mass or upper mass limit of the dust necessary for reproducing the 
SED observed at 200 days is denoted by $M_{2,j}$, where Si grains are 
excluded because their condensation time could be much later than 200 
days (see text).
References for the optical constants used in the calculation are shown 
in the last column. The optical constants of MgSiO$_3$ at $\lambda\leq 
0.3~\mu$m are replaced by that of Mg$_2$SiO$_4$ 
(see Hirashita et al. 2008).}
\tablerefs{
(1) Edo (1983); 
(2) Toon et al. (1976);
(3) Dorschner et al. (1995);
(4) Semenov et al. (2003);
(5) Philipp (1985); 
(6) Roessler \& Huffman (1991);
(7) Piller (1985)
}
\end{deluxetable}


\newpage

\begin{thebibliography}{}

\bibitem[Anupama et al.(2008)]{anu08} 
    Anupama, G. C., et al. 2008, in preparation
\bibitem[Anupama et al.(2001)]{anu01}
    Anupama, G. C., Sivarani, T., \& Pandey, G. 2001, \aap, 367, 506
\bibitem[Arkharov et al.(2001)]{ark06} 
    Arkharov, A., Efimova, N., Leoni, R., Di Paola, A., Di Carlo, E., 
    \& Dolci, M. 2006, Atel, 961, 1
\bibitem[Bianchi \& Schneider(2007)]{bia07} 
    Bianchi, S., \& Schneider, R. 2007, \mnras, 378, 973
\bibitem[Clayton(1979)]{cla79} 
    Clayton, D. D. 1979, \apss, 65, 179
\bibitem[Colgan et al.(1994)]{col94} 
    Colgan, S. W. J., Haas, M. R., Erickson, E. F., Lord, S. D., \&
    Hollenbach, D. J. 1994, \apj, 427, 874 
\bibitem[Di Carlo et al.(2007)]{dic07} 
    Di Carlo, E., et al. 2007, \apj, accepted (astro-ph/0712.3855)
\bibitem[Dorschner et al.(1995)]{dor95}
    Dorschner, J., Begemann, B., Henning, Th., Jaeger, C., \&
    Mutschke, H. 1995, \aap, 300, 503
\bibitem[Douvion et al.(2001)]{dou01} 
    Douvion, T., Lagage, P. O., \& Pantin, E. 
    2001, \aap, 369, 589
\bibitem[Edo(1983)]{edo83} 
    Edo, O. 1983, PhD Dissertation, Dept. of Physics, University of Arizona
\bibitem[Elmhamdi et al.(2004)]{elm04} 
    Elmhamdi, A., Danziger, I. J., Cappellaro, E., Della Valle, M., 
    Gouiffes, C., Phillips, M. M., \& Turatto, M.,
    2004, \aap, 426, 963
\bibitem[Elmhamdi et al.(2003)]{elm03} 
    Elmhamdi, A., Danziger, I. J., Chugai, N., Pastorello, A., Turatto, M.,
    Cappellaro, E., Altavilla, G., Benetti, S., Patat, F., \& Salvo, M. 
    2003, \mnras, 338, 939
\bibitem[Ennis et al.(2006)]{eni06} 
    Ennis, J. A., Rudnick, L., Reach, W. T., Smith, J. D., Rho, J., 
    Delaney, T., Gomez, H., \& Kozasa, T. 2006, \apj, 652, 376
\bibitem[Fassia et al.(2000)]{fas00} 
    Fassia, A., et al., 2000, \mnras, 318, 1093 
\bibitem[Fassia et al.(2001)]{fas01} 
    Fassia, A., et al., 2001, \mnras, 325, 907 
\bibitem[Foley et al.(2007)]{fol07} 
    Foley, R. J., Smith, N., Ganeshalingam, M., Li, W., Chornock, R., 
    \& Filippenko, A. V. 2007, \apj, 657, L105  
\bibitem[Fransson et al.(2005)]{fra05} 
    Fransson, C., et al., 2005, \apj, 622, 991 
\bibitem[Gerardy et al.(2000)]{ger00} 
    Gerardy, C. L., Fesen, R. A., H\"{o}frich, P., \& Wheeler, J. C. 
    2000, \apj, 119, 2968  
\bibitem[Hendry et al.(2005)]{hen05} 
    Hendry, M. A., et al. 2005, \mnras, 359, 906
\bibitem[Hirashita et al.(2008)]{hir08} 
    Hirashita, H., Nozawa, T., Kozasa, T., \& Takeuchi, T. T.
    2008, \mnras, 384, 1725
\bibitem[Immler et al.(2008)]{imm08} 
    Immler, S., et al. 2008, \apj, 674, L85
\bibitem[Kawabata et al.(2008)]{anu08} 
    Kawabata, K. S., et al. 2008, in preparation
\bibitem[Kozasa et al.(1996)]{koz96} 
    Kozasa, T., Dorschner, J., Henning, Th. \& Stognienko, R. 
    1996, \aap, 307, 551
\bibitem[Kozasa et al.(1989)]{koz89} 
    Kozasa, T., Hasegawa, H., \& Nomoto, K. 1989, \apj, 344, 325
\bibitem[Kozasa et al.(1991)]{koz91} 
    Kozasa, T., Hasegawa, H., \& Nomoto, K. 1991, \aap, 249, 474
\bibitem[Landau & Pitaevski(1981)]{lan81}
    Landau, E. M., \& Pitaevski, L. P. 1981, Physical Kinetics 
    (Pergamon, Oxford)
\bibitem[Liu et al.(2000)]{liu00} 
    Liu, Q.-Z., Hu, J.-Y., Hang, H.-R., Qiu, Y.-L., Zhu, Z.-X., \& 
    Qiao, Q.-Y. 2000, \aaps, 144, 219
\bibitem[Lucy et al.(1989)]{luc89} 
    Lucy, L. B., Danziger, I. J., Gouiffes, C., \& Bouchet, P.  1989, 
    in IAU Colloq. 120, Structure and Dynamics of Interstellar Medium, 
    ed. G. Tenorio-Tagle, M. Moles, \& J. Melnick
    (LNP 350; Berlin: Springer), 164
\bibitem[Maeda et al.(2007)]{mae07} 
    Maeda, K., Tanaka. M., Nomoto, K., Tominaga, N., Kawabata, K.,
    Mazzali, P. A., Umeda, H., Suzuki, T., \& Hattori, T.
    2007, \apj, 666, 1069
\bibitem[Mattila et al.(2008)]{mat08} 
    Mattila, S., et al. 2008, \mnras, submitted (astro-ph/0803.2145)
\bibitem[Meikle et al.(2007)]{mei07} 
    Meikle, W. P. S., et al. 2007, \apj, 665, 608
\bibitem[Meikle et al.(1993)]{mei93} 
    Meikle, W. P. S., Spyromilio, J., Allen, D. A., Varani, G.-F., \&
    Cumming, R. J. 1993, \mnras, 261, 535
\bibitem[Minezaki et al.(2007)]{min07} 
    Minezaki, T., Yoshii. Y., \& Nomoto, K. 2007, IAU Circ., 8833, 2
\bibitem[Nakano et al.(2006)]{nak06} 
    Nakano, S., Itagaki, K., Puckett, T., \& Gorelli, R. 
    2006, CBET, 666, 1 
\bibitem[Nozawa et al.(2006)]{noz06} 
    Nozawa, T., Kozasa, T., \& Habe, A. 2006, \apj, 648, 435
\bibitem[Nozawa et al.(2003)]{noz03b} 
    Nozawa, T., Kozasa, T., Umeda, H., Maeda, K., \& Nomoto, K.
    2003, \apj, 598, 785
\bibitem[Pastorello et al.(2007)]{pas07} 
    Pastorello, A., et al. 2007, \nat, 447, 829
\bibitem[Philipp (1985)]{phi85}
    Philipp, H. R. 1985, in
    Handbook of Optical Constants of Solids, ed. E. D. Palik,
    Academic Press, San Diego, p. 719
\bibitem[Piller(1985)]{pil85}
    Piller, H. 1985, in
    Handbook of Optical Constants of Solids, ed. E. D. Palik,
    Academic Press, San Diego, p. 571
\bibitem[Pooley et al.(2002)]{poo02} 
    Pooley, D., et al., 2002, \apj, 572, 932 
\bibitem[Pozzo et al.(2004)]{poz04} 
    Pozzo, M., Miekle, W. P. S., Fassia, A., Geballe, T., Lundqvist, P.,
    Chugai, N. N., \& Sollerman, J. 2004, \mnras, 352, 457
\bibitem[Rho et al.(2007)]{rho07}
    Rho, J., Kozasa, T., Smith, J., Rudnick, L., Ennis, J.,
    Reach, W., DeLaney, T., \& Gomez, H. 2008, \apj, 673, 271
\bibitem[Roessler \& Huffman(1991)]{roe91}
    Roessler, D. M., \& Huffman, D. R. 1991, in Handbook of
    Optical Constants of Solids II, ed. E. D. Palik, Academic
    Press, San Diego, p. 919
\bibitem[Sakon et al.(2007)]{sak07} 
    Sakon, I., et al. 2007, \apj, submitted (astro-ph/0711.4801)
\bibitem[Semenov et al.(2003)]{sem03}
    Semenov, D., Henning, Th., Helling, Ch., Ilgner, M., \& Sedlmayr, E.
    2003, \aap, 410, 611
\bibitem[Shigeyama et al.(1988)]{shi88}
    Shigeyama, T., Nomoto, K., \& Hashimoto, M.
    1988, \aap, 196, 141
\bibitem[Smith et al. (2008)]{smi08} 
    Smith, N., Foley, R. J., \& Filippenko, A. V. 
    2008, \apj, accepted (astro-ph/0704.2249)
\bibitem[Sugerman et al.(2006)]{sug06} 
    Sugerman, B. E. K., et al. 2006, Science, 313, 196
\bibitem[Todini \& Ferrara(2001)]{tod01} 
    Todini, P., \& Ferrara, A. 2001, \mnras, 325, 726
\bibitem[Tominaga et al.(2005)]{tom05} 
    Tominaga, N., et al. 2005, \apj, 633, L97
\bibitem[Tominaga et al.(2007)]{tom07} 
    Tominaga, N., et al. 2007, \apj, submitted (astro-ph/0711.4782)
\bibitem[Toon, Pollack, \& Khare(1976)]{too76}
    Toon, O. B., Pollack, J. B., \& Khare, B. N. 
    1976, J. Geophys. Res., 81, 5733
\bibitem[Umeda \& Nomoto(2002)]{ume02} 
    Umeda, H., \& Nomoto, K. 2002, \apj, 565, 385
\bibitem[Whitelock et al.(1989)]{whi89} 
    Whitelock, P. A., et al.  1989, \mnras, 240, 7P
\bibitem[Wooden et al.(1993)]{woo93} 
    Wooden, D. H., Rank, D. M., Bregman, J. D., Witteborn, F. C.,
    Tielens, A. G. G. M., Cohen, M., Pinto, P. A., \& Axelrod, T. S.  
    1993, \apjs, 88, 477
\bibitem[Woosley(1988)]{woo88}
    Woosley, S. E. 1988, \apj, 330, 218
\end{thebibliography}
\end{document}